\documentclass[ reprint,
notitlepage, amsmath,amssymb,
 aps,
pre, 
]{revtex4-2}

\renewcommand{\exp}[1]{\mathrm{e}^{#1}}
\usepackage{mathtools}
\usepackage{graphicx,color}
\usepackage{dcolumn}
\usepackage{bm}
\usepackage{hyperref}

\usepackage{nicefrac}

\let\vec\bm
\newcommand{\grad}{\vec{\nabla}}

\let\epsilon\varepsilon

\date{October 2023}

\begin{document}
\title{Comment on ``Steady-state distributions and nonsteady dynamics in nonequilibrium systems''}
\author{Horst-Holger Boltz}
\email[]{horst-holger.boltz@uni-greifswald.de}

\author{Thomas Ihle}

\affiliation{Institute for Physics, University of Greifswald, 
17489 Greifswald, Germany}

\begin{abstract}
We comment on a work by T. Liverpool (Phys. Rev. E {\bf 101} 042107 (2020)). We show that a theorem stated in that work is not correct. While this is inconsequential to the bulk of the work (and holds asymptotically as we show), it is instructive to highlight the importance of phase space compression in active systems with generalized dynamics.
\end{abstract}
\maketitle

\citeauthor{liverpool} approached the steady-state distributions of non-equilibrium systems with a focus on the physics of active matter systems.~\cite{liverpool} We want to comment on aspects of this interesting work. The problem is formulated as a somewhat generic dynamical system
\begin{align}
    \frac{\mathrm{d}\vec x}{\mathrm{d}t} = - \mathsf{D}\cdot \grad \mathcal{H}+\vec w + \vec \xi \text{.} \label{eq:dyn}
\end{align}

The time evolution of $\vec x$ thus has three principal components: a gradient flow with a corresponding scalar function $\vec H$ (modulated by a mobility matrix $\mathsf{D}$), a flow field, $\vec w$, that cannot be attributed to a gradient flow (breaking detailed balance) and a Gaussian noise with
\begin{align}
    \langle \xi_i(t) \rangle = 0, \quad
    \langle \xi_i(t)\xi_j(t')\rangle = 2 \theta D_{ij} \delta(t-t') \text{,}
\end{align}
wherein $\theta>0$ represents the strength of the noise. Having effective dynamics and interactions such as those arising in the modeling of active matter in mind, there are barely any constraints to the flows in \eqref{eq:dyn}. For this comment, we restrict ourselves to the case of trivial mobilities, i.e. $D_{ij}=\delta_{ij}$.

The Fokker-Planck equation corresponding to the Langevin equation \eqref{eq:dyn} is
\begin{align}
    \partial_t P &= \sum \grad_i (-F_i P + \theta\grad_i P) \label{eq:fp}
\end{align}
wherein we introduced the short-hand $F_i=-\grad_i\mathcal{H}+w_i$. Stationary solutions of \eqref{eq:fp} correspond to steady-state distributions of the process \eqref{eq:dyn}. In ref.~\citenum{liverpool}, theorems are stated with respect to such stationary solutions. \emph{Theorem 1a} considers the trajectories $\vec x(t)=\vec X(t)$ with
\begin{align}
\dot{\vec X}(t) &= \vec F + \theta \grad h \label{eq:typ}
\end{align}
where $h$ is the Cole-Hopf transform of the stationary distribution, i.e. $P_\text{steady}\propto \exp{-h}$ with $\partial_t P_\text{steady}=0$. From \eqref{eq:fp}, it follows that $h$ is a solution to
\begin{align}
    0&=\sum_i \left[ \theta (\grad_i h)^2 + (\grad_i h) F_i - \theta \grad_i^2h -\grad_i F_i \right] \label{eq:heq} \text{.}
\end{align}
The solutions of eq.~\eqref{eq:typ} are labeled {\it typical trajectories} that, together with fluctuations around them, would constitute {\it generalized steady states}. By construction they have the property that the stationary solutions to the Fokker-Planck equation \eqref{eq:fp} also solve the Liouville equation for this deterministic system. 

It is claimed that $\rho(\vec x)=P_\text{steady}(\vec x(t))$ is constant along the trajectories that are solutions of \eqref{eq:typ}. This statement is not correct. The change along a trajectory is given by
\begin{align}
\frac{\mathrm{d}P}{\mathrm{d}t}(\vec x,t) &= \dot{\vec{x}} \cdot \grad P + \partial_t P \text{.} \label{eq:dPdt}
\end{align}
For the stationary distribution $\rho(\vec X)$ this gives
\begin{align}
    \frac{\mathrm{d}\rho}{\mathrm{d}t}(\vec X) &= \dot{\vec X} \cdot \grad \rho=(\vec F + \theta \grad h) \cdot (- \grad h) \rho\\
    &\stackrel{\eqref{eq:heq}}{=} - \rho \sum_i \left(\theta \grad_i^2 h + \grad_i F_i \right) \text{.} \label{eq:changerho}
\end{align}
In the absence of the non-gradient flows $\vec w$, i.e. $F_i=-\grad_i \mathcal{H}$, the steady-state solution is given by Boltzmann weights $h=\mathcal{H}/\theta$ and the terms in \eqref{eq:changerho} do indeed cancel out. For clarity, we note that this is analog to but not identical with Liouville's theorem as \eqref{eq:dyn} is not a Hamiltonian flow, in absence of a symplectic structure. For truly active systems with general interactions, however, there is generally no analogon to Liouville's theorem: The flow in phase space is compressible. In the language of kinetic theory, this amounts to a change of the superposition principle~\cite{kreuzer} by a phase compression factor, see for example refs.~\citenum{ihle,ihle2}.  Notably, the contributions in \eqref{eq:dyn} can always be decomposed such that the non-gradient part is solenoidal, i.e. $\vec F = - \grad \mathcal{H}_\text{sol} + \vec w_\text{sol}$ with $\grad\cdot\vec w_\text{sol}=0$, see for example ref.~\citenum{gloetzl}. However, $P_\text{sol}\propto \exp{-\mathcal{H}_\text{sol}/\theta}$ is only a steady solution, cp. eq.~\eqref{eq:heq}, if $\grad \mathcal{H}_\text{sol} \cdot \vec w_\text{sol}=0$. One example for such a system is the noisy Hopf system considered in ref.~\citenum{liverpool}. In general, this perpendicularity condition would require the additional flow $\vec w$ to drive the system on submanifolds of constant (generalized) energy (i.e. $\mathcal{H}$ is an integral of motion for $\vec X$). On the other hand, the stationary measure in a noisy gradient-flow-system can be sampled by means of an appropriately chosen non-reversible (active) flow, which is the rationale underlying non-reversible Monte Carlo techniques, see for example ref.~\citenum{michel}. 

We note that the rate of change in \eqref{eq:changerho} is identical to the quantity $\mathcal{C}$ (or $-\mathcal{C}$ as the sign is not consistent along the two definitions) introduced in ref.~\citenum{liverpool}, which in generic cases is not a constant in phase-space.

\begin{figure}
    \centering
    \includegraphics[width=\linewidth]{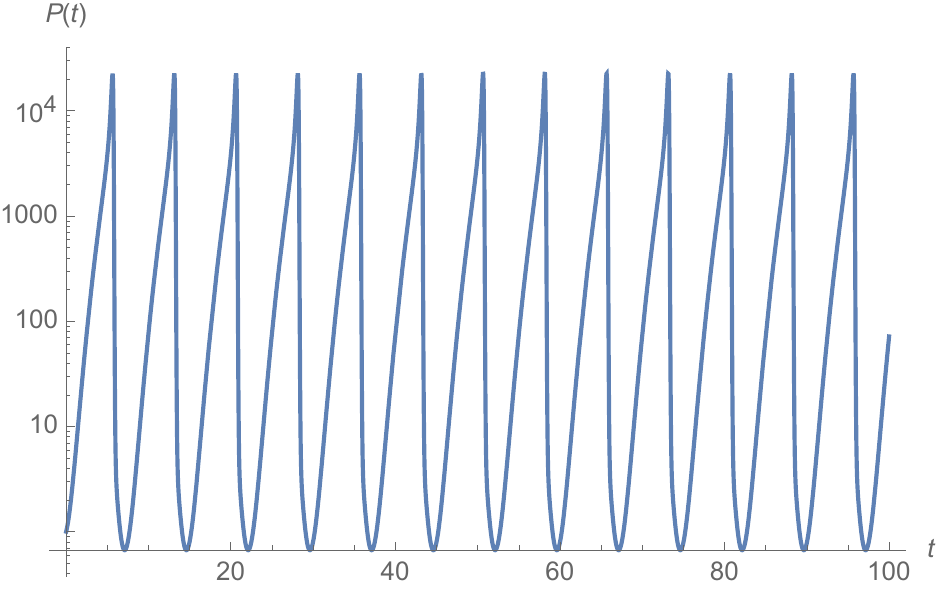}
    \caption{Time evolution of $P\propto \exp{-h(\vec X(t))}$. Highlighting that the probability is not conserved along the trajectory, but varies. Here, we use the approximate expansion solution from ref.~\citenum{liverpool}, the ratio to a numerically found empirical distribution (cp. FIG.~\ref{fig:figh}) of which varies by orders of magnitudes along the typical trajectory. This is a shortcoming of the expansion order, but the actual distribution is also not constant along any meaningful trajectory.}
    \label{fig:fig2}
\end{figure}

\begin{figure*}
    \centering
    \includegraphics[width=\linewidth]{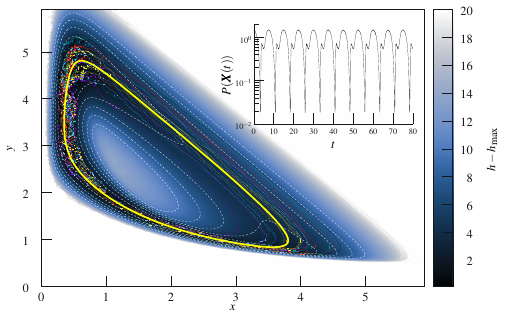}
    \caption{Cole-Hopf transform $h$ of the empirically found stationary measure, $P_\text{steady}\propto \exp{-h}$, for the Brusselator system described in the main text. We added (dashed, white) equipotential lines as visual aids. The image visually corroborates the finding of the text and FIG.~\ref{fig:fig2}, that in this system there are no orbits around the high-energy peak in the middle along which $P_\text{steady}$ is constant that could be considered typical. The suggested typical trajectory of Liverpool is shown in yellow, some exemplary trajectory as dotted lines around it. The bundle of trajectories is tightly focused on parts of the orbit and rather dispersed on others.  The distribution was found from direct integration of the equations of motion \eqref{eq:brusselator} and comprises of roughly $10^{13}$ time steps in the steady state. We highlight the fact that $P$ is not constant along the suggested $\vec X$ trajectory in the inset (scaling such that the shown quantity is effectively $h$). The relative spread in the steady-state probability along the suggested trajectory \eqref{eq:typ} is around two orders of magnitude in this system. From the picture alone, it is also evident that this would still occur if the orbit followed $-\grad h$ and, thus, was located directly in the canyon of the $h$-landscape.}
    \label{fig:figh}
\end{figure*}

\begin{figure*}
    \centering
    \includegraphics[width=\linewidth]{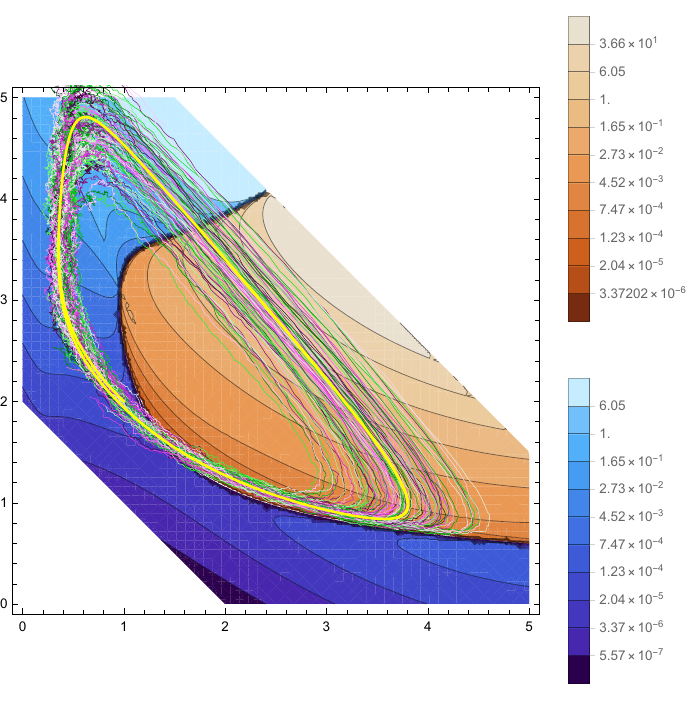}
    \caption{Color map of the change in probability along a trajectory, $\frac{\mathrm{d}P}{\mathrm{d}s} = \frac{\mathrm{d}P}{\mathrm{d}t} / \lvert \dot{ \vec{x}}\rvert $ based on the approximate stationary solution for the Brusselator with $\mu=1$, $\lambda=3$ and $\theta=1/2 (0.1)^2$ given in ref~\citenum{liverpool}. The color-coding is done on a logarithmic scale, with reddish colors corresponding to negative values and blueish values to positive values. The thick yellow line corresponds to the typical trajectory $\vec X$. The discussed effect is also apparent by considering an ensemble (here $N=100$) of trajectories (colored in green to white to pink, for visual clarity). The bundle of trajectories is tightly focused on parts of the orbit and rather dispersed on others.}
    \label{fig:fig1}
\end{figure*}

We make this explicit for the proposed case of the Brusselator as considered in ref.~\citenum{liverpool}. In this two-dimensional system, the dynamics of $\vec x=(x,y)^\mathrm{T}$ are given by \begin{align}\dot{\vec x}&=F+\vec \xi, \vec F = (\mu + x^2y - (\lambda+1)x, \lambda x -x^2y)^\mathrm{T}.\label{eq:brusselator}\end{align} One possible decomposition could be $\mathcal{H}_\text{sol}=-x^3y/3+x^4/12+(\lambda+1)/2\, x^2$ and $\vec w_\text{sol}=(\mu+x^3/3, \lambda x - x^3/3 -x^2y)^\mathrm{T}$ and there exists no decomposition with polynomial flow fields that allows for $\mathcal{H}_\text{sol}=\theta h$. For the system with $\mu=1$, $\lambda=3$ and $\theta=(0.1)^2/2$, an approximate solution for $h$ is given in ref.~\citenum{liverpool}. We have checked that this solution is indeed a solution to the stationary Fokker-Planck equation to the stated order (albeit neither stable nor accurate for large arguments, which is irrelevant here).  As we show in FIG.~\ref{fig:fig2}, the typical trajectories found from integrating eq.~\eqref{eq:typ} for this stationary distribution do not correspond to constant values of $h$. We illustrate this further by depicting the change in the value of the stationary distribution along a trajectory for both the deterministic typical trajectory $\vec X$ as well as an ensemble of stochastic trajectories, see FIG.~\ref{fig:fig1}. We present $h$ as inferred from direct numerical integration of the stochastic equations of motion, eq.~\eqref{eq:brusselator}, in FIG.~\ref{fig:figh}. 

However, we want to make note that \emph{Theorem 1a} of ref.~\citenum{liverpool} is \emph{asymptotically} correct, if the system can be meaningfully expanded around vanishing noise, $\theta\to 0$. In this {\it weak noise} case, the stationary distribution can be written as $P_\text{steady}\propto \exp{-g/\theta}$ with $g=g_0 + \theta g_1 +\mathcal{O}(\theta^2)$. Using this Ansatz and separating terms by orders of $\theta$, the Fokker-Planck equation can be reduced to an effective Hamilton-Jacobi equation for $g_0$ and the so-called transport equation for $g_1$, see for example ref.~\citenum{gaspard} for an instructive discussion. Their respective time evolutions are given by
\begin{align}
    \partial_t g_0 &= - \vec F \cdot \grad g_0 - (\grad g_0)^2 \label{eq:g0}\\
    \partial_t g_1 &= - \grad\cdot \vec F -2 (\grad g_0)\cdot (\grad g_1) - \vec F \cdot \grad g_1 + \Delta g_0 \label{eq:g1}
\end{align}
The former equation is the Hamilton-Jacobi equation for a Hamiltonian system with the so-called Freidlin-Wentzell Hamilton~\cite{freidlinwentzell} function $\mathcal{H}_\text{FW}=\vec p^2 + \vec F \cdot \vec p$ wherein the canonical moments are given by $p_i = \frac{1}{2} (\dot x_i - F_i)$. The subspace of interest is $\vec p=0$. As pointed out before,~\cite{gaspard} the solution to the latter equation, the transport equation, ultimately gives rise to a weak-noise propagator whose amplitude is modulated by \begin{align}\exp{-\int_0^t \mathrm{Tr}(\frac{\partial^2 \mathcal{H}_\text{FW}}{\partial \vec x\partial \vec p}(t'))\mathrm{d}t'}.\end{align} In the relevant subspace, the integrand becomes
\begin{align}
\left.    \mathrm{Tr}(\frac{\partial^2 \mathcal{H}_\text{FW}}{\partial \vec x\partial \vec p})\right\rvert_{\vec p=0} =\grad\cdot \vec F\end{align} rephrasing the point that the divergence of the generalized forces corresponds to phase space compression. Going back to the approach of ref.~\citenum{liverpool}, we can consider trajectories $\vec X=\vec X_0 + \theta \vec X_1+\mathcal{O}(\theta^2)$. From the stationary versions of eqs. \eqref{eq:g0} and \eqref{eq:dPdt}, we directly find that 
\begin{align}
0&= \dot {\vec X}_0 - \vec F - \grad g_0 \\
0&= \dot {\vec X}_0 \grad g_1 + \dot{\vec X}_1 \text{.}
\end{align}
Thus, the trajectories that are considered in ref.~\citenum{liverpool} are actually the zeroth order contribution. Before contemplating this further, it is also interesting to explicitly state the next order. Using \eqref{eq:g1}, we can find
\begin{align}
    0&=\grad \cdot \vec F + (\dot{\vec X}_1 -\grad g_1)\grad g_0 + \Delta g_0;
\end{align}
one reasonable solution is $\dot{ \vec{X}}_1 = \grad g_1 + q \grad g_0$ with \begin{align}q = - \frac{\grad\cdot \vec F + \Delta g_0}{(\grad g_0)^2}.\end{align} This is far from unique as locally any vector field orthogonal to $\grad g_0$ could be added.  In a general setting, the limit $\theta\to 0$ is peculiar with respect to stationary distributions. In a chaotic system with a non-hyperbolic attractor (which is the generic case for an interacting system with a useful stationary state), the existence of a stationary probability (that does not depend on the initial state) is directly bound to finiteness of the noise, i.e $\theta>0$.~\cite{anishchenko2005}

Another situation in which the theorem does hold, are systems in which the steady state is such that all the detailed balance breaking fluxes in the non-equilibrium steady state ({\em NESS}) are trivial. One class of such systems is composed of systems that have a global flocking states. There are two types of generalized forces here, that could give rise to change along the typical trajectory: the propulsion and the alignment interactions. In a flocking state, the alignment interaction can effectively be of a gradient-type in a co-moving frame as changes in neighborhood topology can be negligible. If the propulsion speed then does not vary spatially or based on orientation (which it could), the original statement is correct.

While the proposed ``typical'' trajectories of eq.~\eqref{eq:dyn} do convey some information in the limit of weak noise as they locally represent the trajectory between points that maximizes the transition probability among all trajectories in the weak-noise approximation, phase space (de-)compression is possible and arguably typical in active matter systems with generalized interactions with one consequence being that, in general, the {\em theorem 1a} of ref.~\citenum{liverpool} does not hold. In fact, in most non-trivial systems there will be no meaningful orbits along which the steady-state probability is constant, potentially limiting the usefulness of the proposed concept of
{\em generalized steady states}.

\bibliography{lit.bib}

\end{document}